\documentclass[a4paper,fleqn]{cas-dc}

\usepackage[numbers,sort&compress]{natbib}
\usepackage{multicol}
\usepackage{caption}
\usepackage{amsmath}
\usepackage{multicol}
\usepackage{fancyhdr}
\usepackage{soul} 
\usepackage{color, xcolor} 
\usepackage{bm}
\soulregister\citep7
\soulregister\ref7

\def\tsc#1{\csdef{#1}{\textsc{\lowercase{#1}}\xspace}}
\tsc{WGM}
\tsc{QE}

\begin{document}

\captionsetup[figure]{name={Fig.}}
\begin{sloppypar}
\let\WriteBookmarks\relax
\def\floatpagepagefraction{1}
\def\textpagefraction{.001}
\let\printorcid\relax 



\title[mode = title]{CloudBrain-MRS: An Intelligent Cloud Computing Platform for in vivo Magnetic Resonance Spectroscopy Preprocessing, Quantification, and Analysis}   

\author[1]{Xiaodie Chen}[style=chinese]
\fnmark[1] 
\author[1]{Jiayu Li}[style=chinese]
\fnmark[1] 
\author[1]{Dicheng Chen}[style=chinese]
\author[1]{Yirong Zhou}[style=chinese]
\author[1]{Zhangren Tu}[style=chinese]
\author[2]{Meijin Lin}[style=chinese]
\author[3]{Taishan Kang}[style=chinese]
\author[3]{Jianzhong Lin}[style=chinese]
\author[4]{Tao Gong}[style=chinese]
\author[5]{Liuhong Zhu}[style=chinese]
\author[5]{Jianjun Zhou}[style=chinese]
\author[6]{Lin Ou-yang}[style=chinese]
\author[7]{Jiefeng Guo}[style=chinese]
\author[1]{Jiyang Dong}[style=chinese]
\author[8]{Di Guo}[style=chinese]
\author[1]{Xiaobo Qu}[style=chinese]
\corref{cor1}






\address[1]{ Department of Electronic Science, Fujian Provincial Key Laboratory of Plasma and Magnetic 
Resonance, Xiamen University, Xiamen, China}
\address[2]{ Department of Applied Marine Physics \& Engineering, Xiamen University, Xiamen, China}
\address[3]{Department of Radiology, Zhongshan Hospital Affiliated to Xiamen University, Xiamen, China}
\address[4]{Departments of Radiology, Shandong Provincial Hospital Affiliated to Shandong First Medical University, Jinan, Shandong, China}
\address[5]{Department of Radiology, Zhongshan Hospital (Xiamen), Fudan University, Xiamen, China}
\address[6]{Department of Medical Imaging of Southeast Hospital, Medical College of Xiamen University, Xiamen, China}
\address[7]{Department of Microelectronics and Integrated Circuit, Xiamen University, Xiamen, China}
\address[8]{School of Computer and Information Engineering, Xiamen University of Technology, 
Xiamen, China}

\cortext[cor1]{Corresponding author.} 

\begin{abstract}
Magnetic resonance spectroscopy (MRS) is an important clinical imaging method for diagnosis of diseases. MRS spectrum is used to observe the signal intensity of metabolites or further infer their concentrations. Although the magnetic resonance vendors commonly provide basic functions of spectra plots and metabolite quantification, the widespread clinical research of MRS is still limited due to the lack of easy-to-use processing software or platform. To address this issue, we have developed CloudBrain-MRS, a cloud-based online platform that provides powerful hardware and advanced algorithms. The platform can be accessed simply through a web browser, without the need of any program installation on the user side. CloudBrain-MRS also integrates the classic LCModel and advanced artificial intelligence algorithms and supports batch preprocessing, quantification, and analysis of MRS data from different vendors. Additionally, the platform offers useful functions: 1) Automatically statistical analysis to find biomarkers for diseases; 2) Consistency verification between the classic and artificial intelligence quantification algorithms; 3) Colorful three-dimensional visualization for easy observation of individual metabolite spectrum. Last, both healthy and mild cognitive impairment patient data are used to demonstrate the functions of the platform. To the best of our knowledge, this is the first cloud computing platform for in vivo MRS with artificial intelligence processing. We have shared our cloud platform at MRSHub, providing free access and service for two years. Please visit \href{https://mrshub.org/software_all/#CloudBrain-MRS}{https://mrshub.org/software\_all/\#CloudBrain-MRS} or \href{https://csrc.xmu.edu.cn/CloudBrain.html}{https://csrc.xmu.edu.cn/CloudBrain.html}. 
\end{abstract}




\maketitle
\let\thefootnote\relax\footnotetext{$^1$Xiaodie Chen and Jiayu Li are co-first authors and contributed equally to this study.}
\let\thefootnote\relax\footnotetext{E-mail addresses: \url{quxiaobo@xmu.edu.cn} (Xiaobo Qu).}
\thispagestyle{empty}
\section{Introduction}
\label{sec:introduction}

Magnetic resonance spectroscopy (MRS) is a non-invasive technique used to quantify metabolites in the human brain to diagnose various diseases, such as breast cancer, craniopharyngiomas, and Rett syndrome \citep{malhi2002magnetic, behrens2007computer, gokcay2002proton, sener2001proton}. However, the acquired MRS signals typically require data preprocessing and quantitative analysis to obtain accurate metabolite concentrations \citep{near2021preprocessing}. The purpose of preprocessing is to reduce the impact of unwanted factors on data quality, such as field inhomogeneities, scanner frequency drift, noise, and subject motion \citep{drost2002proton,near2021preprocessing}.
Preprocessing steps may include coil merging, lineshape correction, denoising, phase correction, frequency alignment, and water suppression \citep{poullet2008mrs,mandal2012vivo,SR}. Quantifying MRS signals is a challenge due to low signal-to-noise ratios (SNR) and  overlapping peaks \citep{AMARES}. Traditional quantification methods  mainly included simple peak integration \citep{jansen20061h} and peak fitting \citep{VARPRO,AMARES,LCModel,QUEST,AQSES} methods. In recent years, the rapid development of deep learning has led to the emergence of new artificial intelligence algorithms in the field of MRS signal preprocessing \citep{preprocessing1,preprocessing2,ReLSTM,jang2021unsupervised} and quantification \citep{quantization1,quantization2,quantization3,lee2019intact, chen2023magnetic}. Then, artificial intelligence software and clinical validation of these new approaches are eagerly needed.
\begin{figure*}[t]
    \centering    \includegraphics[width=0.9\textwidth,height=0.41\textwidth]{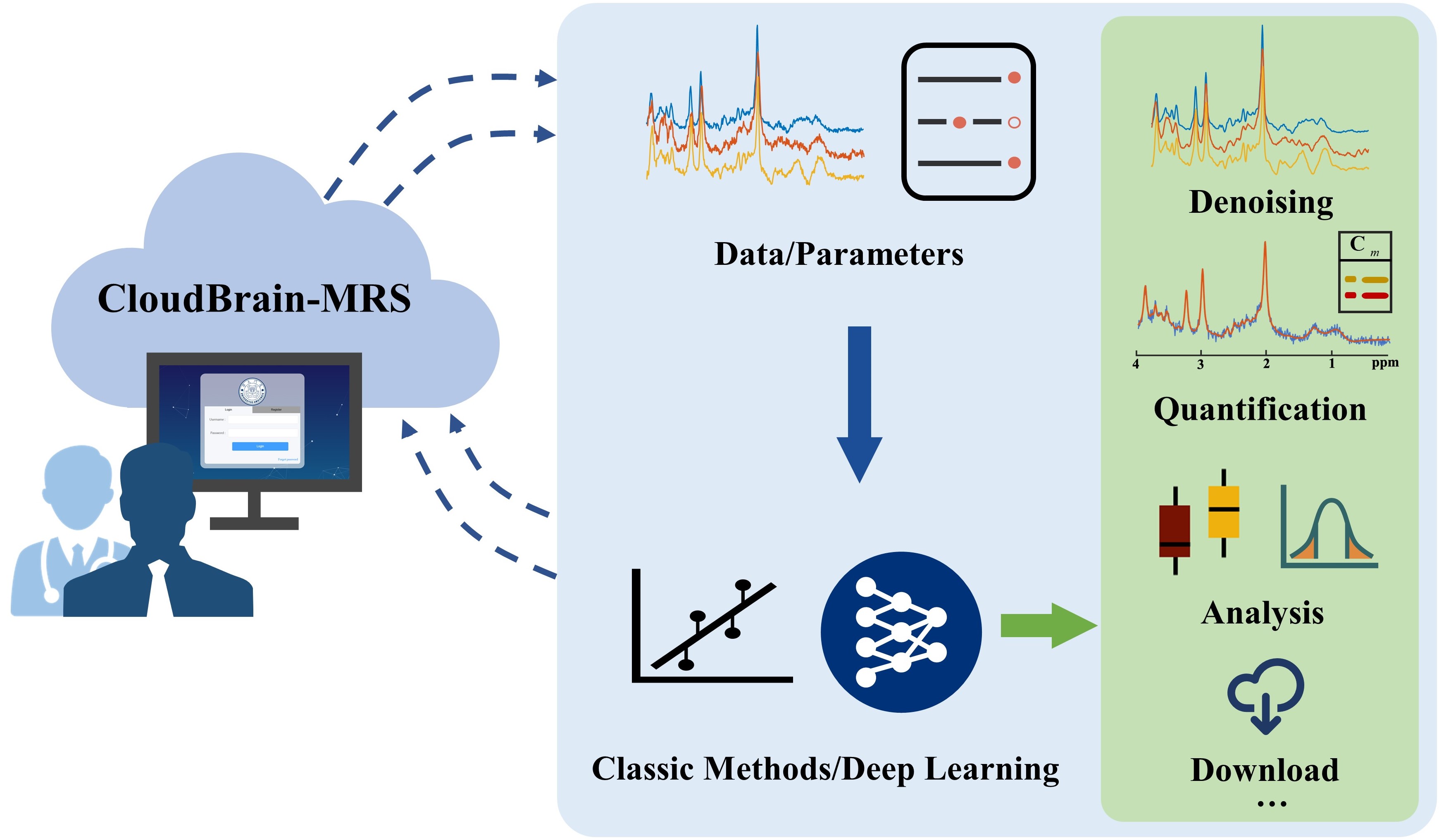}
    \caption{CloudBrain-MRS. $\mathbf C_{m}$ indicates the concentration of metabolites.}
    \label{FIG1}
\end{figure*}

\begin{table}[t]
 \caption{Programming languages of some tools for MRS.}\label{tbl1}
 \setlength{\tabcolsep}{11mm}{
 \begin{tabular}{lc}
  \toprule
  Tool & Language \\
  \midrule
 LCModel \citep{LCModel}& Fortran\\
 JMRUI \citep{JMRUI1,JMRUI2}& Java\\
 TARQUIN \citep{TARQUIN}& C++\\
 FSL-MRS \citep{FSL-MRS}& Python\\
 Osprey \citep{Osprey}& MATLAB\\
 Gannet \citep{Gannet}& MATLAB\\
 FID-A \citep{FID-A}& MATLAB \\
  \bottomrule
 \end{tabular}}
\end{table}

Currently, there are various open-source tools available for preprocessing, quantification, and analysis of MRS signals, including LCModel \citep{LCModel}, JMRUI \citep{JMRUI1,JMRUI2}, TARQUIN \citep{TARQUIN}, FSL-MRS \citep{FSL-MRS}, Osprey \citep{Osprey}, FID-A \citep{FID-A}, and Gannet \citep{Gannet}, as shown in Table \ref{tbl1}. But none of them implements deep learning. LCModel is a widely used tool written in Fortran for MRS quantification, but users need to compile and install it on a Linux PC, which requires certain skills. JMRUI \citep{JMRUI1,JMRUI2} provides a user-friendly graphical interface based on the Java framework, but users need to install Java to run it. TARQUIN \citep{TARQUIN} is a GUI-based tool that relies on a C++ library, but batch processing requires command line program. FSL-MRS \citep{FSL-MRS} is a collection of Python modules that require users to install environment dependencies. Osprey \citep{Osprey} and FID-A \citep{FID-A} are fully integrated MRS data analysis pipeline that relies on the MATLAB development environment. Gannet \citep{Gannet} is a tool for the automated quantification of edited MRS data and is run via MATLAB commands.  While these tools provide a user-friendly interface, they still require users to compile source code, download dependencies, or install the software. Furthermore, none of these tools include deep learning algorithms, which is a significant limitation for the current research in the era of artificial intelligence. There is a strong need for a user-friendly system that can enable biomedical researchers and clinical radiologists to apply these advanced algorithms effectively to clinical research. 

In the past few decades, there have been several MRS cloud platforms for simulating basis sets \citep{hui2022mrscloud}, reconstructing spectrum in accelerated nuclear magnetic resonance \citep{wang2022sparse}, and integrating MRI into the radiation therapy planning workflows \citep{gurbani2019brain}, Not limited to MRS, cloud platforms have been applied in magnetic resonance imaging (MRI), including reconstructing and evaluating images in fast imaging \citep{SPM,CloudBrain-ReconAI,xue2015distributed}, processing and analyzing images \citep{mori2016mricloud, milletari2018cloud}, and simulating MRI signals \citep{xanthis2019coremri}. Cloud computing provides an easily accessible, flexible and scalable platform. Users need not worry about hardware maintenance and management, and thus, they can focus on innovation and core tasks. In this paper, we present a full suite of processing and postprocessing MRS on cloud computing platform.

In this study, we propose CloudBrain-MRS, a cloud-based platform for automated data preprocessing, quantification, and analysis of MRS data, as in Fig. \ref{FIG1}. The platform provides both hardware and software, and the latter includes advanced deep learning  denoising method ReLSTM \citep{ReLSTM} and quantification method QNet \citep{chen2023magnetic}, and the mainstream quantification tool LCModel. Users can batch preprocess and quantify MRS data online via a browser without any coding or installation of the development environment. The platform also includes a statistical analysis module to evaluate biomarker differences in quantification results between health and patient groups, and a series of visualization services to help users evaluate these results. The platform has developed as well a consistency analysis module for users to evaluate the reliability of the platform's quantification algorithms compared with LCModel.

CloudBrain-MRS has several significant advantages over the existing tools for MRS. Firstly, it can be run using only a browser, eliminating the need for powerful hardware configurations and client installation. Secondly, it greatly reduces the requirements for the technical skill of users. Thirdly, developers can expand and maintain the system, with updates being delivered simultaneously to clients. Fourthly, this platform has been developed with a module for the statistical analysis of biomarkers. Such a function is very important for clinical research but not provided by the previous MRS tools (Table \ref{tbl1}). Lastly, as far as we know, CloudBrain-MRS is the first cloud-based computational platform that applies artificial intelligence to in vivo MRS.

\begin{figure}[t]
    \centering
    \includegraphics[width=0.45\textwidth,height=0.48\textwidth]{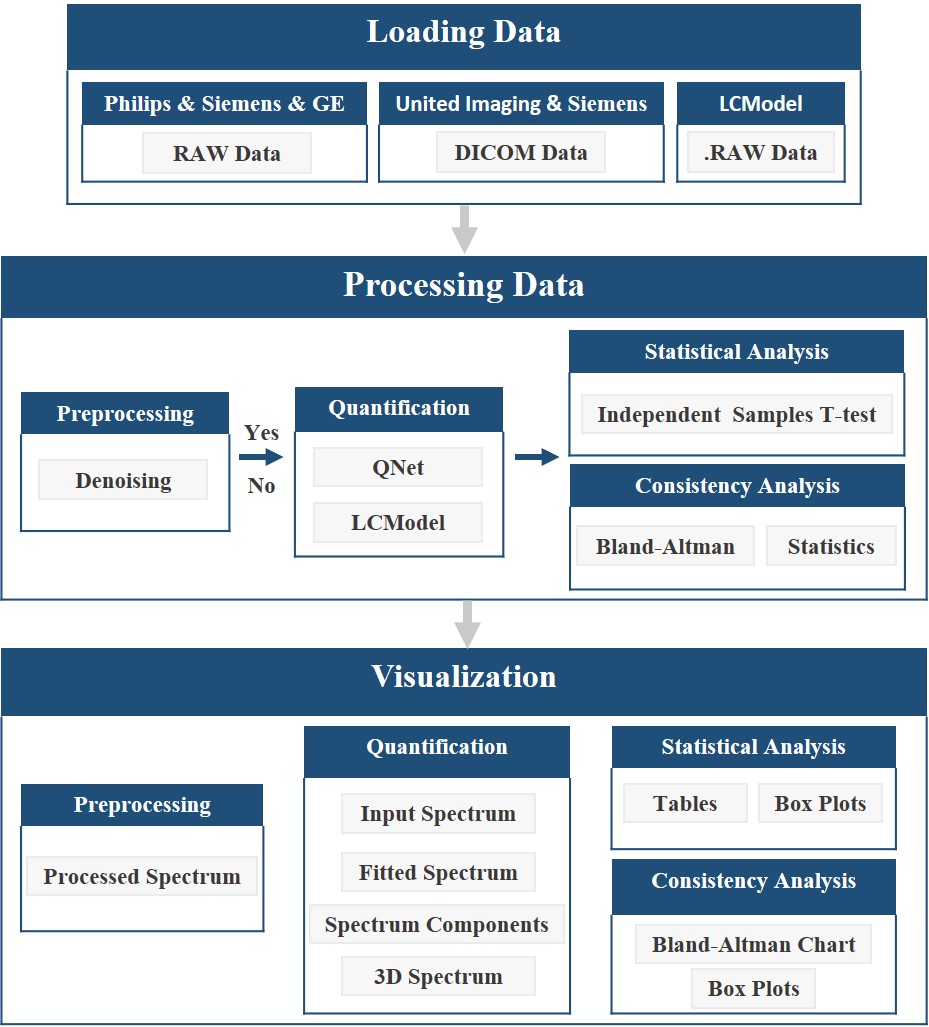}
    \caption{The whole workflow of CloudBrain-MRS.}
    \label{FIG2}
\end{figure}

\section{Platform design and implementation}
\label{Platform design and implementation}
\subsection{Workflow summary}
\label{Workflow summary}
Currently, the platform mainly contains two functional modules: Intelligent quantification and automatic analysis. Users can register an account or use our demo account (username: demo\_csg, password: csg12345678!). The manual on the homepage also can help users to get started quickly. The workflow of CloudBrain-MRS is illustrated in Fig. \ref{FIG2} and can be described in detail as follows:

(1) Load the data and the corresponding parameters. The platform currently supports reading RAW data from Philips, Siemens, and GE, DICOM data from United Imaging and Siemens, and also supports LCModel's data format.

(2) Invoke the quantification model and quantify the data in batch or single, and save the quantification results. If user chooses to preprocess the data, denoising will be performed before quantification.

(3) Generate four types of visual spectra based on the quantification results: Inputted spectrum, fitted spectra of overall and individual metabolites, and 3D visualized spectrum. If denoising is applied, the spectra before and after denoising will be shown.

(4) Extract the quantitative results for analysis and generate the corresponding analysis charts. Statistical analysis will generate box plots and trilinear tables. The function of ‘‘Consistency Analysis’’ will generate Bland-Altman charts and box plots.

\begin{figure}[t]
    \centering
    \includegraphics[width=0.4\textwidth,height=0.48\textwidth]{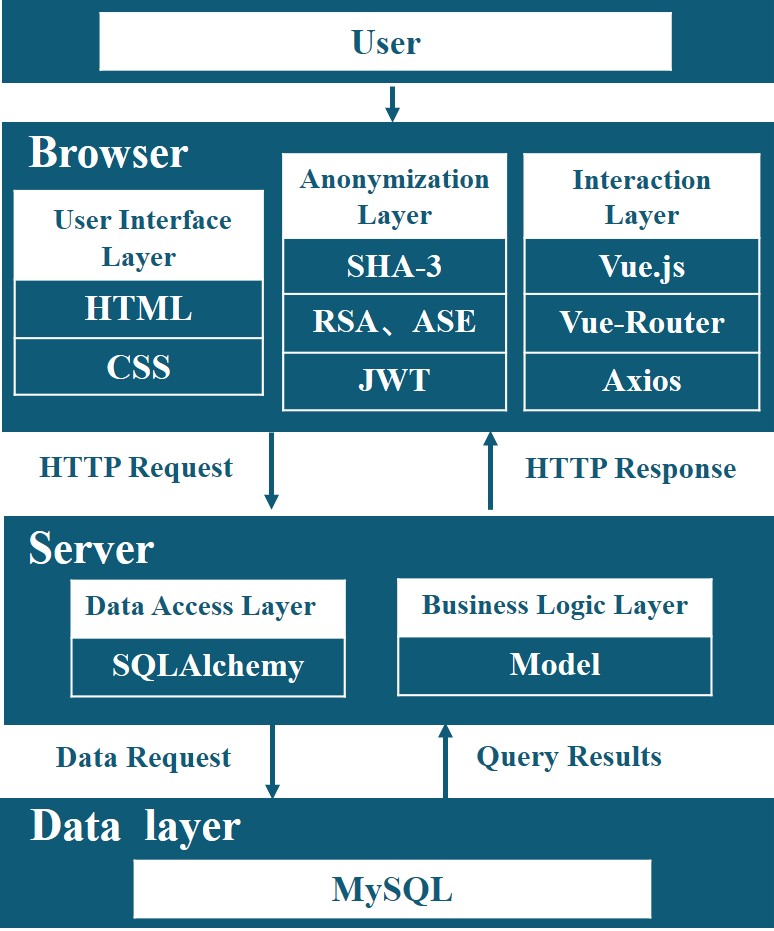}
    \caption{System architecture of CloudBrain-MRS.}
    \label{FIG3}
\end{figure}

\subsection{Architecture of the system}
\label{ System architecture}

To enhance user-friendliness, CloudBrain-MRS adopts the browser/service (B/S) working model, which stands for a browser-request/server-response model. The system architecture can be divided into three parts, namely the browser, server, and database. Fig. \ref{FIG3} displays the interactions between each part and the libraries they depend on. Below are the functions and dependencies of each part:

(1) The browser is used for user interaction with the server, which includes a user interface layer, an anonymization layer, and an interaction layer. The user interface layer utilizes hypertext markup language (HTML) and cascading style sheets (CSS) technologies to display data and operations to the user, providing a user-friendly interface and interaction experience. Meanwhile, the interaction layer is implemented through the Vue framework, which sends the user's request to the server and processes the response through the hypertext transfer protocol (HTTP). The anonymization layer is for data security and privacy protection and the details will be described in Section \ref{ System security}.

(2) The server is responsible for processing data. After receiving HTTP requests from the browser, the server uses SQLAlchemy to automatically generates structured query language (SQL) statements to interact with the data layer. Once the data has been retrieved, the server processes it based on the pre-existing business logic. This is done through the business logic layer, which encapsulates the application model and executes the application policy. This enables the system to perform various functions such as data processing, calculation, and analysis. 

(3) The data layer stores all data, including user information, quantitative results, statistical analysis results, and other relevant data, in MySQL. 

In summary, the server of our platform can be considered a machine with hardware and algorithms. The server is equipped with an Intel Xeon Processor with 4 cores, 62GB of RAM, and an NVIDIA Tesla K40M GPU, which will be utilized to accelerate the training and testing of the deep learning models. The developer of our platform is responsible for hosting and maintaining the server, and the user can access and use the service directly through a browser only. The platform sets a storage space limit of 1GB per registered user. With a user concurrency of 100, the throughput is 152.8 transactions per second. Using the B/S mode can reduce the cost for users and improve the scalability of the system.

\begin{figure*}[t]
    \centering
    \includegraphics[width=0.55\textwidth,height=0.7\textwidth]{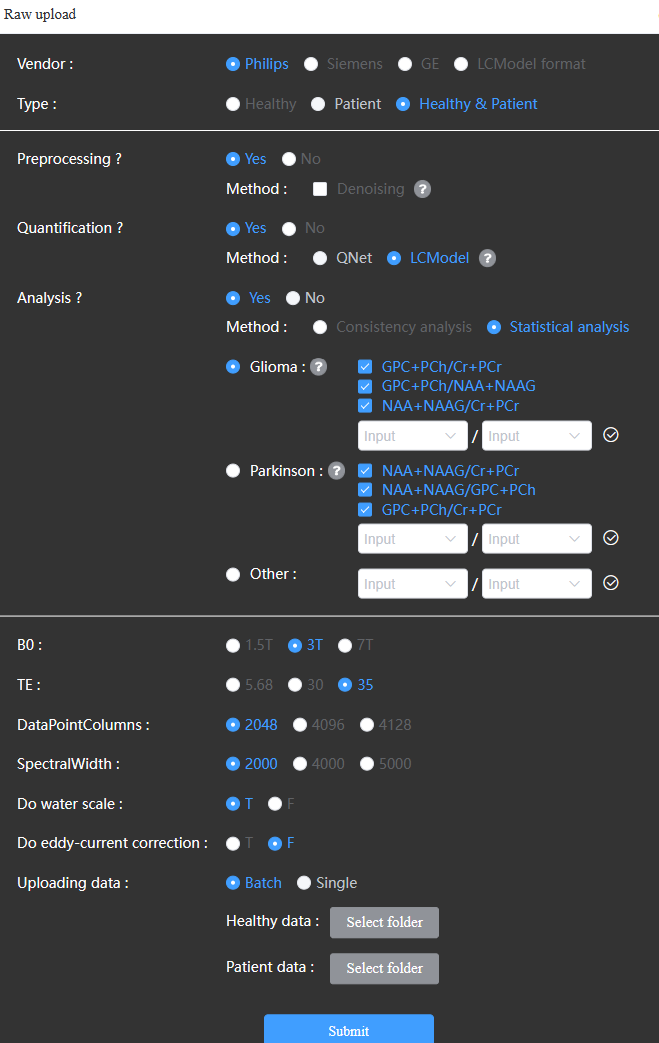}
    \caption{ The user interface of CloudBrain-MRS for RAW data. The user interface will change based on user choices.}
    \label{FIG4}
\end{figure*}

\subsection{Security and privacy of the system}
\label{ System security}
In the cloud system, uploaded files are desensitized, and sensitive information such as names are deleted, but the information needed for data analysis such as age and gender is retained. Patient privacy is handled at the browser and no patient-identifiable information is transmitted to our server. Users have the right to delete data. Once deleted, both the original and processed data are permanently deleted on the server. 

For data secure transmission, CloudBrain-MRS adopts measures such as encrypted transmission and identity authentication to prevent sensitive information from being illegally obtained. The platform uses the 2048-bit Rivest-Shamir-Adleman (RSA) algorithm to encrypt sensitive information before transmission. JSON Web Token (JWT) is utilized for validating the user's login status, with the Advanced Encryption Standard (ASE) algorithm ensuring the JWT is transmitted with encryption for secure authentication. 

For data storage security, the platform sets up a white list of allowed ports to impose strict access restrictions on the database and adds protection against distributed denial of service (DDOS) attacks. Data in the database and cache are stored using encrypted storage.

\section{Algorithms in the system}
\label{Method}
\subsection{Signal model}
\label{Signal model}
CloudBrain-MRS models the MRS signal as a combination of the metabolite signal, background signal from MacroMolecules (MMs), and noise \citep{LCModel}. The metabolite signal is a linear combination of the elements in the basis set. The complex signal data $Y[(n\Delta t)|\varepsilon _Y]$ \citep{LCModel} can be modeled as:
\begin{equation} \label{eq1} 
    \begin{aligned}
&\hat{Y}[(n\Delta t)|\varepsilon _Y] = \text{exp}[-i(\varphi _0+n\Delta t\varphi _1)][B(n\Delta t)
\\&+\sum_{l=1}^{N_M}C_lM_l(n\Delta t;\gamma_l,f_l)]+\varepsilon _Y(n\Delta t),
n=0,1,2,...,N-1,
    \end{aligned}
\end{equation}
where $N$ is the length of the signal, $\Delta t$ is the sampling interval, $\varepsilon _Y(n\Delta t)$ is the complex white Gaussian noise, and $\sigma(Y[n\Delta t]|\varepsilon _Y)$ is the standard deviation of the noise. $B(n\Delta t)$ denotes the background signal. $\varphi _0$ and $\varphi _1$ denote the zero-order and first-order phases due to non-ideal acquisition conditions. $C_l$ denotes the $l$-th concentration factor, and $N_M$ is the number of metabolites. $M_l(n\Delta t;\gamma_l,f_l)$ is the signal of the $l$-th metabolite modeled in the basis set and is disturbed by imperfection factors (IFs) \citep{LCModel} as follows:
\begin{equation}
    \begin{aligned}
    M_l(n\Delta _t;\gamma_l,f_l)=\mathscr{F}\{m_l(n\Delta t)exp[-(\gamma_l+if_l)\Delta t]\},
    \end{aligned}
\end{equation}
where $\mathscr{F}$ denotes the discrete Fourier transform, $m_l(n\Delta t)$ is the time-domain signal of $M_l(n\Delta _t;0,0)$. $\gamma_l$ and $f_l$ denote the linewidth deviation and frequency drift due to non-ideal conditions.

\subsection{Denoising}
\label{Denoising_method}
For in vivo spectra, low metabolite concentrations and non-ideal conditions can result in low SNR, which leads to difficulty in quantification and analysis \citep{poullet2008mrs,LCModel}. Denoising is a signal preprocessing technique used to remove noise from a signal and improve its quality. 

CloudBrain-MRS has deployed an end-to-end deep learning denoising model called Refusion Long Short-Term Memory (ReLSTM) \citep{ReLSTM} for preprocessing data and improving the SNR  of data (Table \ref{tbl2}). This model has been trained with in vivo brain spectra to map MRS time-domain data with low SNR (24 Signal Averages (SA)) to high SNR (124 SA). 

After obtaining $Y[(n\Delta t)|\varepsilon _Y]$ with a few repeated samples ($\geq$24 SA), the platform applies the denoising model to obtain a high SNR spectrum $\hat{Y}[(n\Delta t)|\varepsilon _{124\text{SA}}]$ close to 124 repeated samples, which further improves the quantitative accuracy of key metabolites \citep{ReLSTM}.
\begin{equation}
    \begin{aligned}
    \hat{Y}[(n\Delta t)|\varepsilon _{124\text{SA}}]=\text{ReLSTM}\{Y[(n\Delta t)|\varepsilon _Y]\}.
    \end{aligned}
\end{equation}

\subsection{Quantification}
\label{Quantification_method}
To ensure the reliability of the quantification results, CloudBrain-MRS integrates LCModel and QNet algorithms.
\subsubsection{LCModel}
\label{LCModel}
LCModel utilizes $N_B$ cubic B-splines $B_j(n\Delta t)$ to model the background signal and utilizes the lineshape coefficients $S_n$ to represent field inhomogeneities, eddy currents, etc. Eq. \ref{eq1} is expressed in LCModel \citep{LCModel} as follows:
\begin{equation} 
    \begin{aligned}
&\hat{Y}[(n\Delta t)|\varepsilon _Y] =  \text{exp}[-i(\varphi _0+n\Delta t\varphi _1)][\sum_{j=1}^{N_B}H_jB_j(n\Delta t)
\\&+\sum_{l=1}^{N_M}C_l\sum_{k=-N_s}^{N_s}S_kM_l(n\Delta t;\gamma_l,f_l)]+\varepsilon _Y(n\Delta t),
    \end{aligned}
\end{equation}
Where $S_k$ and $H_j$ are the lineshape coefficients and the B-spline coefficients, respectively.

The whole optimization problem of the fitting model \citep{LCModel} is defined as:
\begin{equation} \label{eq5}
    \begin{aligned}
    &\mathop{\text{min}}\limits_{\mathbf{C}}\frac{1}{\sigma^2\{Y[(n\Delta t)|\varepsilon _Y]\}}\text{Re}\{Y[(n\Delta t)|\varepsilon _Y]-\hat{Y}[(n\Delta t)|\varepsilon _Y]\}^2\\
&+||\alpha_SR_S\mathbf{S}||^2+||\alpha_BR_B\mathbf{H}||^2+\sum_{l=1}^{N_M}\{\frac{[\gamma_l-\gamma_l^0]^2}{\sigma^2(\gamma_l)}+\frac{f_l^2}{\sigma^2(f_l)}\},
    \end{aligned}
\end{equation}   
where Re\{$\cdot$\} denotes the real part of the complex vector, $\mathbf{S}$ and $\mathbf{H}$ are the vector of $S_k$ and $H_j$, respectively. $R_S$ and $R_H$ are regular matrices with smoothing constraints on $\mathbf{S}$ and $\mathbf{H}$, respectively. $a_S$ and $a_H$ are weighting factors used to balance between the regular terms. The last terms in Eq. \ref{eq5} represent prior normal probability distributions for the parameters $\gamma_l$ and $f_l$, making the solution more stable. LCModel solves Eq. \ref{eq5} with the Levenberg-Marquardt (LM) algorithm and a limited-memory algorithm for bound constrained optimization (L-BFGS-B) to obtain metabolite concentrations $\hat{\mathbf{C}}$ \citep{LCModel}.

Compared to other commonly used quantification methods,  LCModel has excellent model building, quantification accuracy, and noise resistance. Although LCModel has been open-sourced, it requires a certain level of compiling and operational skills for installation and use. To enhance user-friendliness, the platform has integrated LCModel with interactions facilitated through a shell script. This integration enables the platform to efficiently process data in batches using LCModel without users understanding the internal operation details. LCModel requires users to provide a sequence-specific basis set. Our platform contains several commonly used basis sets that can be automatically selected based on vendor and sequence parameters. The platform will provide more types of basis sets for service users in the future.

\subsubsection{QNet}
\label{QNet}
CloudBrain-MRS also has deployed an artificial intelligence quantification model from QNet. The model contains a deep learning network for predicting IFs $\{\hat{\varphi_0},\hat{\gamma},\hat{\mathbf{f}}\}$ to solve nonlinear problems based on the powerful ability of deep learning \citep{chen2023magnetic}, as follows:
\begin{equation} 
    \begin{aligned}
    \{\hat{\varphi_0},\hat{\gamma},\hat{\mathbf{f}}\} = \mathcal{N}_{\text{extraction}}(Y[(n\Delta t)|\varepsilon _Y]|\mathbf{\Theta}_{\text{extraction}}),
    \end{aligned}
\end{equation}   
where $\mathcal{N}_{\text{extraction}}(\cdot)$ is a deep learning network for predicting $\{\hat{\varphi_0},\hat{\gamma},\hat{\mathbf{f}}\}$ from input $Y[(n\Delta t)|\varepsilon _Y]$ with network parameters $\mathbf{\Theta}_{\text{extraction}}$, and $\hat{\mathbf{f}}=[\hat{f_1},\hat{f_2},...,\hat{f}_{N_M}]$. The network consists of 3 stacked convolutional blocks (SCBs) and 2 fully connected layers. Each SCBs consists of 2 convolutional layers and a maximum pooling layer, each convolutional layer is followed by the non-linear activation function Rectified Linear Unit (ReLU).

Since the background signal is more variable for in vivo data and difficult to model accurately, the platform used a large number of simulated background signals to train a deep learning network $\mathcal{N}_{\text{prediction}}(Y[(n\Delta t)|\varepsilon _Y]|\mathbf{\Theta}_{\text{prediction}})$, which can predict $\hat{B}(n\Delta t)$ directly from the input $Y[(n\Delta t)|\varepsilon _Y]$ with the network parameters $\mathbf{\Theta}_{\text{prediction}}$. The module \citep{chen2023magnetic} can be represented as follows:
\begin{equation} 
    \begin{aligned}
    \hat{B}(n\Delta t) = \mathcal{N}_{\text{prediction}}(Y[(n\Delta t)|\varepsilon _Y]|\mathbf{\Theta}_{\text{prediction}}),
    \end{aligned}
\end{equation} 

The network consists of 6 SCBs and 2 fully connected layers. Finally, metabolite concentrations $\hat{\mathbf{C}}$ can be estimated using linear least squares \citep{chen2023magnetic}:
\begin{equation} 
    \begin{aligned}
    &\mathop{\text{min}}\limits_{\mathbf{C}}||\text{exp}[-i(\hat{\varphi_0})][\sum_{l=1}^{N_M}C_lM_l(n\Delta t;\hat{\gamma_l},\hat{f_l})-Y[(n\Delta t)|\varepsilon_Y]\\
    &+\hat{B}(n\Delta t)]||^2.
    \end{aligned}
\end{equation} 

The method combines the interpretability of the magnetic resonance signal model and the nonlinear learning ability of the neural network to achieve fast and accurate quantification of MRS. Experimental results show that QNet has a more stable quantification performance than LCModel at different SNRs \citep{chen2023magnetic}.

\subsection{Statistical analysis}
\label{Statistical analysis method}
Analyzing differences in biomarkers between health and patient groups can help researchers gain a better understanding of changes in the biochemistry of the human body, providing valuable data for disease prevention. For example, in the study of Alzheimer’s disease, N-acetylaspartate (NAA)$/$Creatine (Cr) has been identified as a potential biomarker for brain dysfunction \citep{wang2015magnetic, kherchouche2022attention}. Some existing statistical software can help with statistical analysis, such as SPSS and Excel. However, users need to organize the quantitative results by themselves, which is very time-consuming. 

CloudBrain-MRS has developed a statistical analysis module that automatically quantifies and analyzes data uploaded by users, as shown in Fig. \ref{FIG4}. The module uses LCModel to quantify data and an independent samples t-test or a Mann-Whitney U-test to analyze whether there is a significant difference between healthy individuals and patients. To analyze the metabolic characteristics of gliomas and Parkinson's disease, the platform provides several key metabolite concentrations as references. For glioma patients, the platform provides tCho (Glycerophosphocholine (GPC)$+$Phosphocholine (PCh))$/$tCr (Cr+Phosphocreatine (PCr)), tCho$/$tNAA (NAA+N-acetylaspartylglutamate (NAAG)), and tNAA$/$tCr as indicators \citep{lazen2023comparison, nakae2017prediction, kazda2016advanced}. For Parkinson's disease, the platform provides tNAA$/$tCr, tCho$/$tCr, and tNAA$/$tCho as indicators \citep{mazuel2016proton, flamez2019influence}. In addition, users can add other indicators according to their research needs for analysis. If the values of these indicators between healthy individuals and patients satisfy the assumption of normal distribution, the platform will use an independent samples t-test to analyze. Instead, a Mann-Whitney U test should be applied. The procedure of an independent samples t-test consists of two stages \citep{kim2019statistical}. In the initial stage, Levene's test is applied to further evaluate the assumption of equal variances. If the assumption is met, the Student's t-test is selected. If not, the Welch's t-test is chosen.

\subsection{Consistency analysis}
\label{Consistency analysis method}
To help users verify the reliability of the traditional quantification method LCModel and artificial intelligence quantification method QNet, CloudBrain-MRS provides a consistency analysis service that is currently limited to healthy individuals, as shown in Fig. \ref{FIG4}. This service has two aspects. Firstly, a Bland-Altman analysis is conducted to evaluate the degree of consistency between the two algorithms. This analysis will calculate the difference and mean of the concentrations quantified by the two algorithms and then will be illustrated by a scatter plot. Secondly, box plots of metabolite concentrations are generated based on the normal concentration ranges \citep{lee2019intact,terpstra2016test,tkavc2009vivo,govindaraju2000proton} in healthy individuals to check the distribution of metabolite concentration values. This enables users to know if the quantification results of the two algorithms fall within the normal concentration ranges. Currently, the consistency analysis mainly focuses on tNAA$/$tCr, tCho$/$tCr, Glx (Glutamate (Glu)+Glutamine (Gln))$/$tCr, myo-Inositol (Ins)$/$tCr, and Glutathione (GSH)$/$tCr.

\subsection{High efficiency of the platform workflow}
\label{Advantages of the platform's workflow over individual components}
To demonstrate the advantage of improving work efficiency, we invited three clinical doctors to use the platform for batch quantification and analysis of 5 spectra acquired on a Philips scanner from 5 healthy volunteers. Using CloudBrain-MRS, users only need to upload data and select parameters to obtain results for preprocessing, quantification, and analysis. The average time spent by all doctors to learn the whole workflow including all the above tasks was 7.33 minutes.

In contrast, the existing workflows without using our platform are time-consuming and require users to perform many steps: (1) Parsing data, using appropriate functions based on different vendors' formats. (2) Performing data preprocessing using packaged functions and models. (3) Converting data formats. Some programs can read raw data from different vendors, but using LCModel to quantify MRS data often requires converting data to the .RAW format. (4) Conducting quantification. Users need to upload basis sets or generate them using software, which can be time-consuming. (5) Organizing results from multiple data quantifications into tables. (6) Using tools like SPSS for data analysis. (7) Additionally, if users need to use advanced deep learning models, they also need to train models separately and preprocess data into formats acceptable to the models.

Thus, the key advantage of this platform lies in providing an integrated workflow that covers the entire process including preprocessing, quantification, and analysis. Compared to the existing workflows of using the individual corresponding programs one by one, it greatly saves users’ time and reduces the requirement for specialized skills.
\begin{figure*}[t]
    \centering
    \includegraphics[width=0.95\textwidth,height=0.4\textwidth]{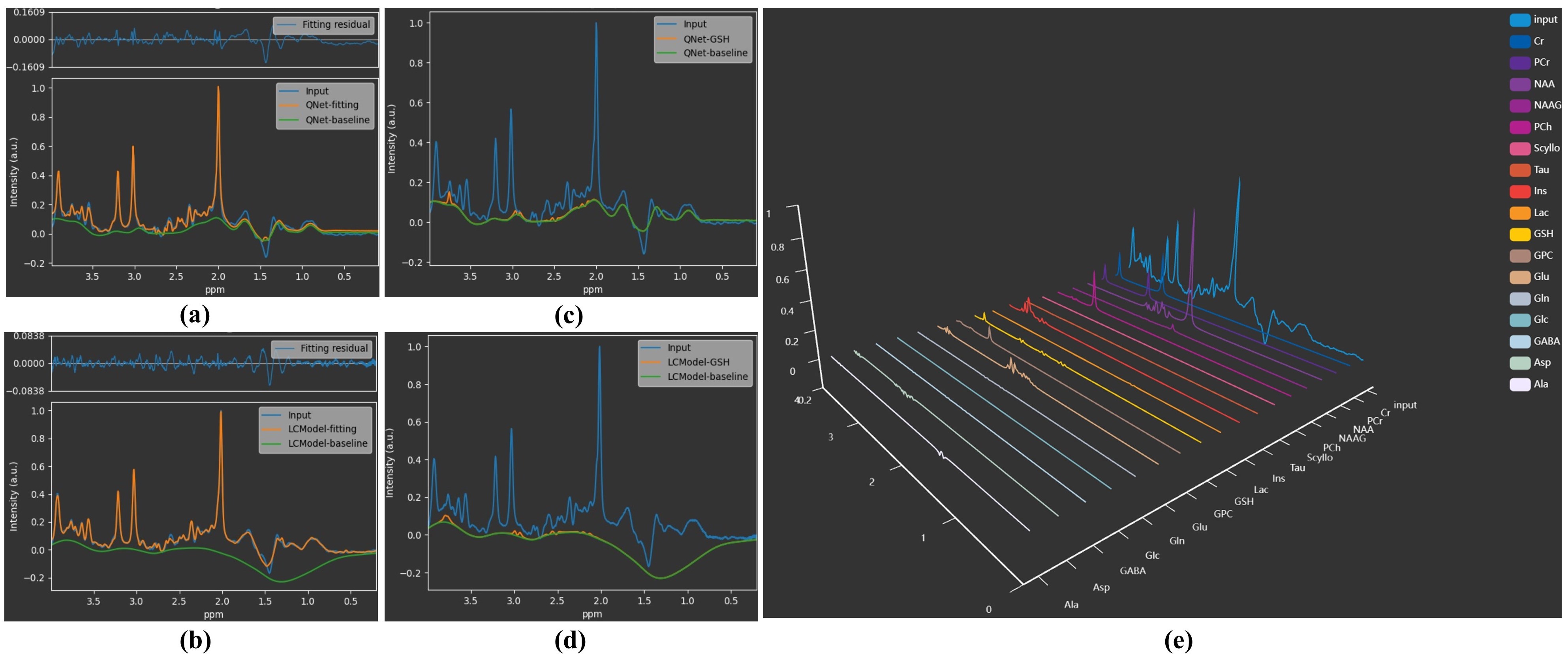}
    \caption{Examples of visualization of CloudBrain-MRS.
(a) and (b) are the fitted spectra of QNet and LCModel, respectively. (c) and (d) are the fitted spectra of GSH with QNet and LCModel, respectively. (e) is the 3D visualization spectra of QNet.}
    \label{FIG5}
\end{figure*}  

\begin{figure}[t]
    \centering
    \includegraphics[width=0.48\textwidth,height=0.3\textwidth]{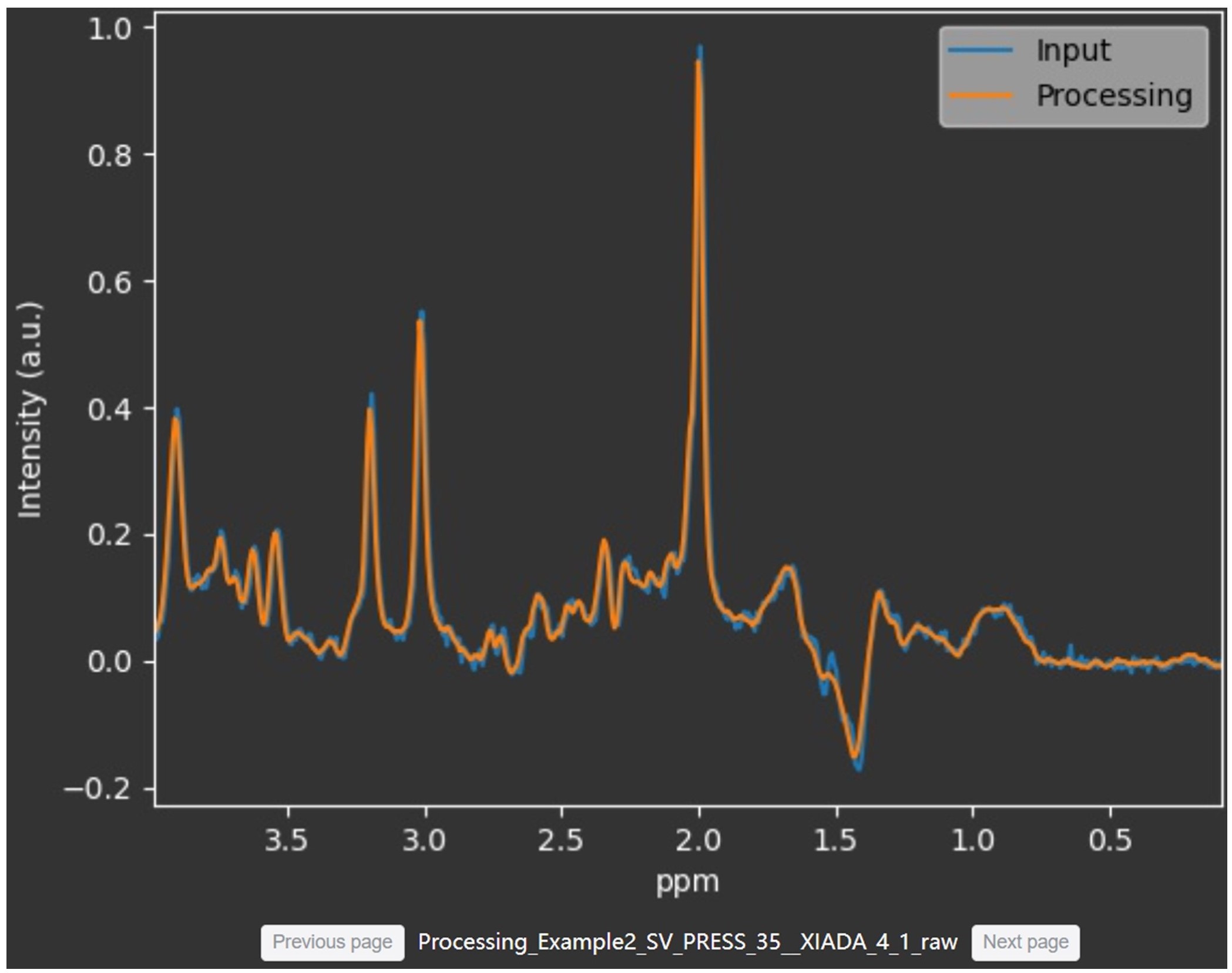}
    \caption{A denoising result of CloudBrain-MRS from a healthy volunteer. The unit of chemical shift is expressed in parts per million (ppm).}
    \label{FIG6}
\end{figure}

\subsection{Visualization}
\label{More visualization}
To help users evaluate quantification results, CloudBrain-MRS has developed a range of visualization tools. Users can view fitted spectra of QNet or LCModel, as shown in Fig. \ref{FIG5}a and Fig. \ref{FIG5}b. Additionally, the platform extracts the fitted results for each metabolite, as shown in Fig. \ref{FIG5}c and Fig. \ref{FIG5}d, which aids in evaluating the contribution of each metabolite. Moreover, the platform utilizes echarts technology to provide 3D visualization of every metabolite, enabling users to have a comprehensive view of the fit results, as in Fig. \ref{FIG5}e.

\section{Demonstrations with in vivo data}
\label{Experiments}
We demonstrate the practicality of CloudBrain-MRS by embedding its services in some simple examples. 

The in vivo data used in Sections \ref{Denoising}, \ref{Quantification}, and \ref{Consistency analysis} were approved by the institutional review board of Xiamen University. A total of 15 single-voxel short-TE PRESS MRS data were collected from 15 healthy volunteers on Philips scanners (3 T field strength, spectral width = 2000 Hz, 2048 points, TR = 2000 ms, TE = 35 ms, voxel size = 20 $\times$ 20 $\times$ 20 mm$^3$, Number of Signal Averages (NSA) = 128). 5 spectra were selected for illustrations in Sections \ref{Denoising} and \ref{Quantification}, and all 15 spectra were used for illustrations in Section \ref{Consistency analysis}.

The in vivo data used in Section \ref{Statistical analysis} were approved by the institutional review board of Shandong Provincial Hospital affiliated to Shandong University. 12 single-voxel short-TE PRESS MRS data were collected from 12 healthy volunteers and 14 single-voxel short-TE PRESS MRS data were collected from 14 mild cognitive impairment (MCI) patients with Philips scanners (3 T field strength, spectral width = 2000 Hz, 2048 points, TR = 2000 ms, TE = 30 ms, voxel size = 20 $\times$ 20 $\times$ 40 mm$^3$, NSA = 128).

\begin{table}[t]
\caption{Improvement of SNR after denoising.}\label{tbl2}
\begin{tabular*}{\tblwidth}{@{}CCC@{}}
\toprule
Test sample No. & SNR before denoising& SNR after denoising\\ 
\midrule
 1& 22& 33\\
 2& 26& 33 \\
 3& 30&  49\\
 4& 34& 46 \\
 5& 25&  42\\
\bottomrule
\end{tabular*}
\end{table}

\subsection{Denoising}
\label{Denoising}
5 spectra (Philips RAW data) from healthy volunteers were uploaded to CloudBrain-MRS for preprocessing tests. The denoising performance is evaluated by the SNR \citep{LCModel}, and the improvement in SNR is summarized in Table \ref{tbl2}. The denoising result demonstrates the effective suppression of noise, as shown in Fig. \ref{FIG6}. Where the blue curve is the input spectrum and the yellow curve is the result after denoising. The name of the spectrum is displayed at the bottom. And users can switch to the results of the previous or next data by clicking on ‘‘Previous page’’ or ‘‘Next page’’ when performing batch processing.

\begin{figure}[t]
    \centering
    \includegraphics[width=0.48\textwidth,height=0.3\textwidth]{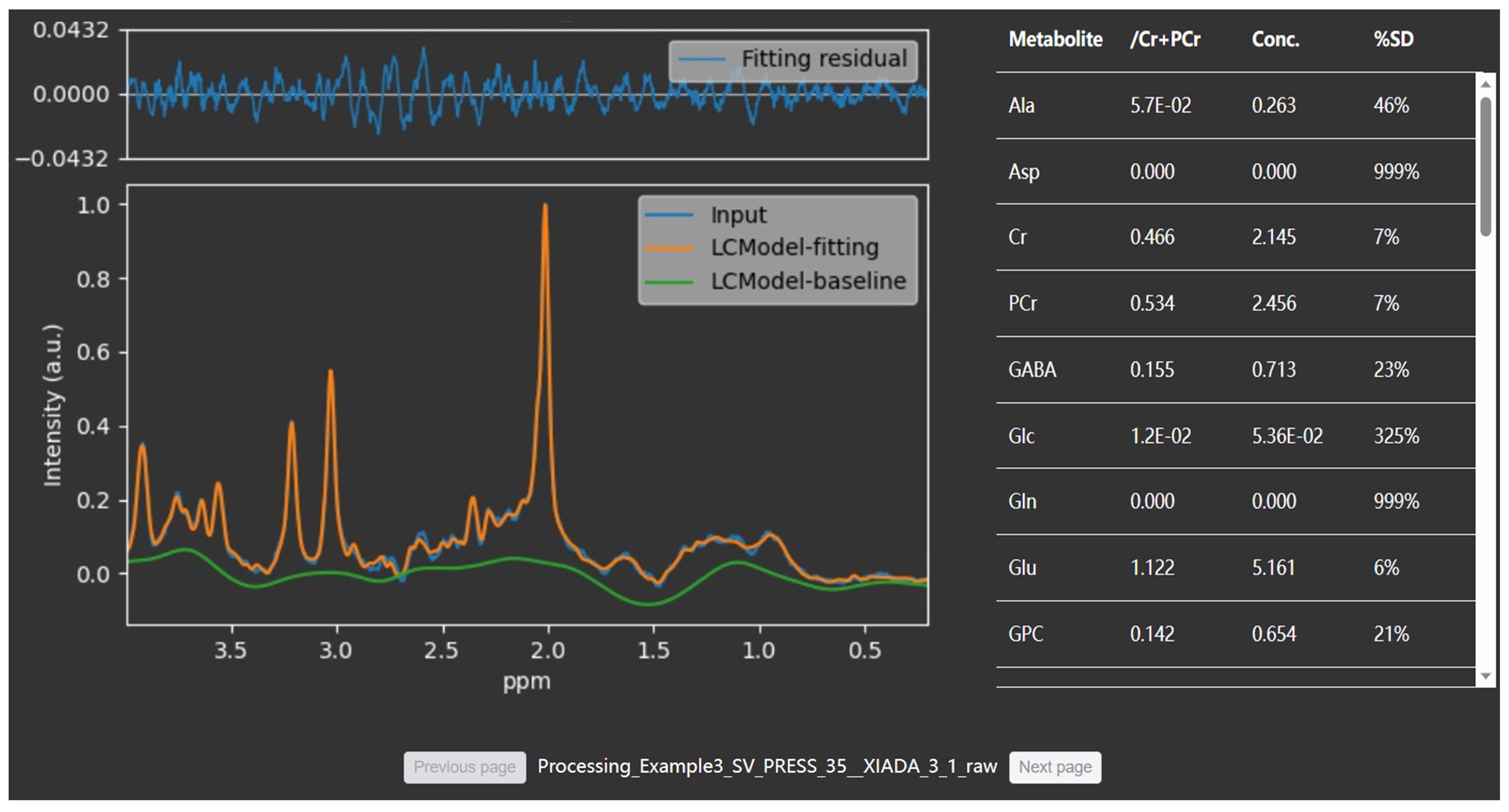}
    \caption{A quantitative result of LCModel from a healthy volunteer. The spectrum  was denoised before quantification.}
    \label{FIG7}
\end{figure}

\begin{figure}[t]
    \centering
    \includegraphics[width=0.48\textwidth,height=0.3\textwidth]{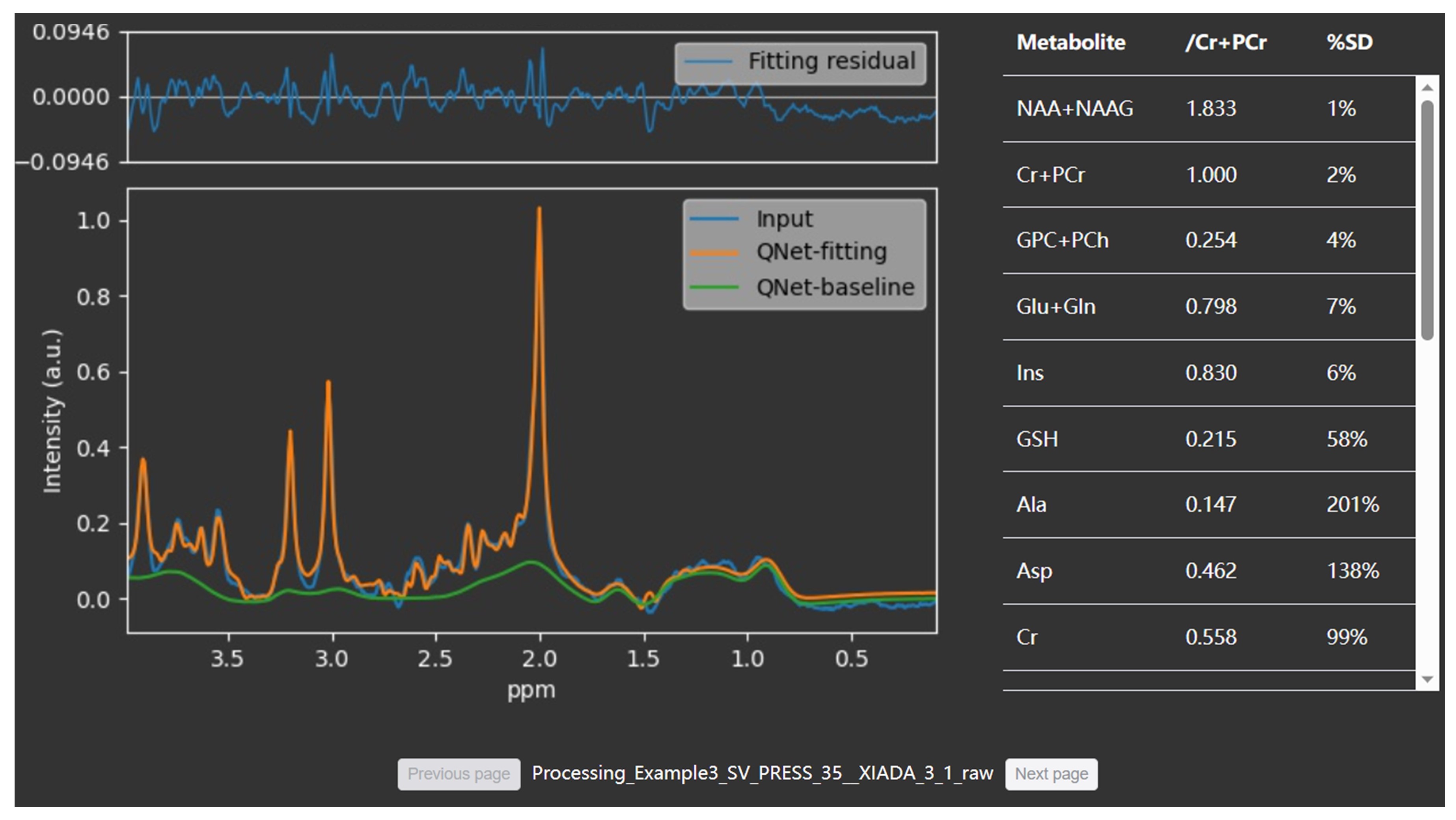}
    \caption{A quantitative result of QNet from a healthy volunteer. The spectrum was denoised before quantification.}
    \label{fig8}
\end{figure}

\begin{figure}[t]
    \centering
    \includegraphics[width=0.48\textwidth]{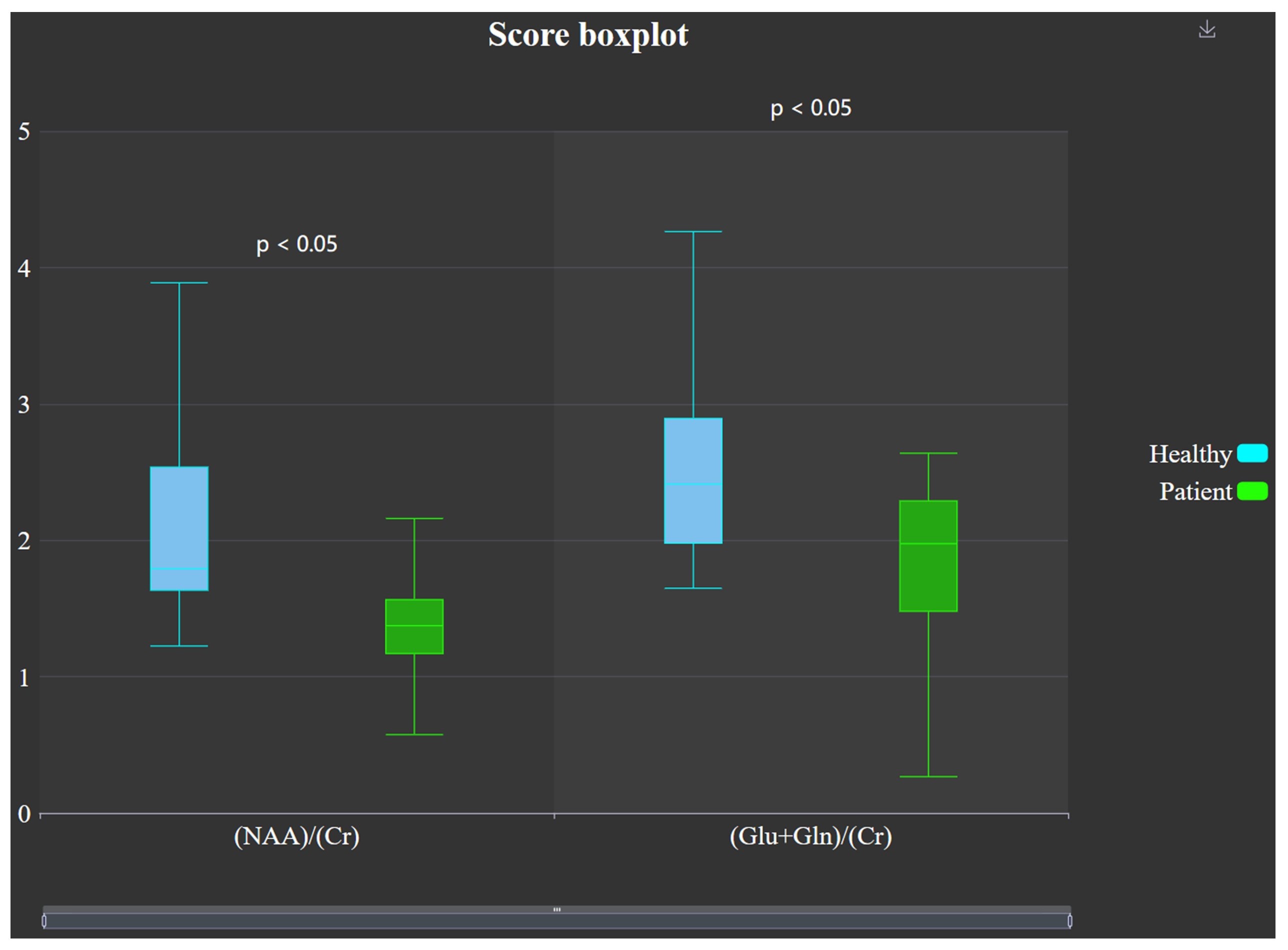}
    \caption{ Box plots of relative metabolite concentrations for statistical analysis between 12 healthy volunteers and 14 MCI patients. A sliding bottom tab bar was designed to view box plots of other indicators. A group with a p-value less than 0.05 will be automatically marked by the platform.}
    \label{FIG9}
\end{figure}

\begin{figure}[t]
    \centering
    \includegraphics[width=0.48\textwidth]{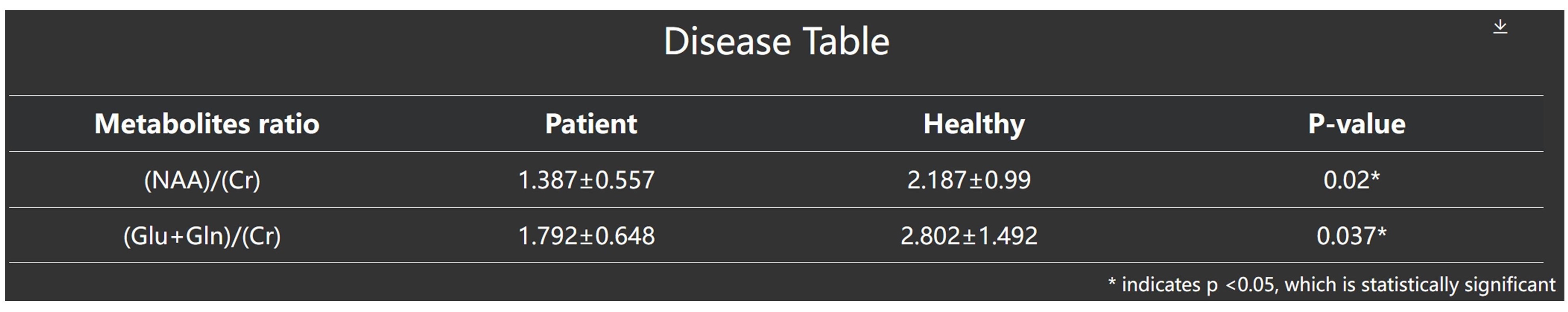}
    \caption{The independent samples t-test results between 12 healthy volunteers and 14 MCI patients. The data are represented as mean $\pm$ standard deviation.}
    \label{FIG10}
\end{figure}

\subsection{Quantification}
\label{Quantification}
The 5 denoised spectra in Section \ref{Denoising} were used to test the two quantification models separately. 

\subsubsection{Quantitative results of LCModel.}
\label{Quantitative results of lcmodel}
One of the LCModel quantification results from the CloudBrain-MRS platform is shown in Fig. \ref{FIG7}. The left bottom of Fig. \ref{FIG7} shows the comparison results between the input spectrum and the LCModel fitted spectrum. The left top of Fig. \ref{FIG7} shows the residual, that is the difference between them. And the right in Fig. \ref{FIG7} presents the quantified concentrations of 17 metabolites. The column of ‘‘Metabolite" is a list of metabolite names, the column of ‘‘$/$Cr$+$PCr’’ indicates the relative concentration of a metabolite to tCr, and ‘‘conc.’’ indicates the absolute concentration. Cramér-Rao Lower Bound (CRLB) is a reliable indicator of minimum errors for estimated parameters \citep{kreis2016trouble}, and ‘‘\%SD’’ represents the CRLB expressed in percent of the estimated concentration \citep{LCModel}. The \%SD ranges from 0 to 999, and a \%SD < 20 is used as a rough criterion of acceptable reliability \citep{LCModel, ReLSTM}. A smaller \%SD value implies a more precise estimate. Different data can be selected by clicking on ‘‘Previous page’’ or ‘‘Next page’’ at the bottom. LCModel takes approximately 6.75 seconds to quantify one spectrum.

\subsubsection{Quantitative results of QNet}
\label{Quantitative results of Qnet}
The result of quantification by QNet for the same spectrum in Section \ref{Quantitative results of lcmodel} is shown in Fig. \ref{fig8}. QNet only provides relative concentrations of metabolites. QNet only takes approximately 5.00 seconds to quantify one spectrum.

\subsection{Statistical analysis}
\label{Statistical analysis}
MCI has been identified as a cognitive disturbance at high risk of dementia \citep{metastasio2006conversion}. MRS can detect biomarkers of MCI for early diagnosis and disease progression \citep{kantarci2013proton}. MCI patients have shown lower levels of NAA$/$Cr and Glx$/$Cr compared with healthy controls \citep{tumati2013magnetic}.

12 spectra (Philips RAW data) from healthy volunteers and 14 spectra from MCI patients were uploaded to CloudBrain-MRS for statistical analysis. Two relative metabolite concentrations, NAA$/$Cr and Glx$/$Cr were selected to evaluate the differences between groups. The results of the independent t-test and box plots generated using the platform are presented in Fig. \ref{FIG9} and  Fig. \ref{FIG10}.  Compared with the healthy controls, MCI patients show decreased levels of NAA$/$Cr (1.387±0.557 for patients, 2.187±0.99 for healthy volunteers) with p $=$ 0.02 and Glx$/$Cr (1.237±0.332 for patients, 0.293±0.038 for healthy volunteers) with p $=$ 0.037. Therefore, it can be concluded that there are statistically significant differences in the two relative metabolite concentrations between groups.
Users can download and save charts of statistical analysis results.

\begin{figure}[t]
    \centering
    \includegraphics[width=0.48\textwidth,height=0.35\textwidth]{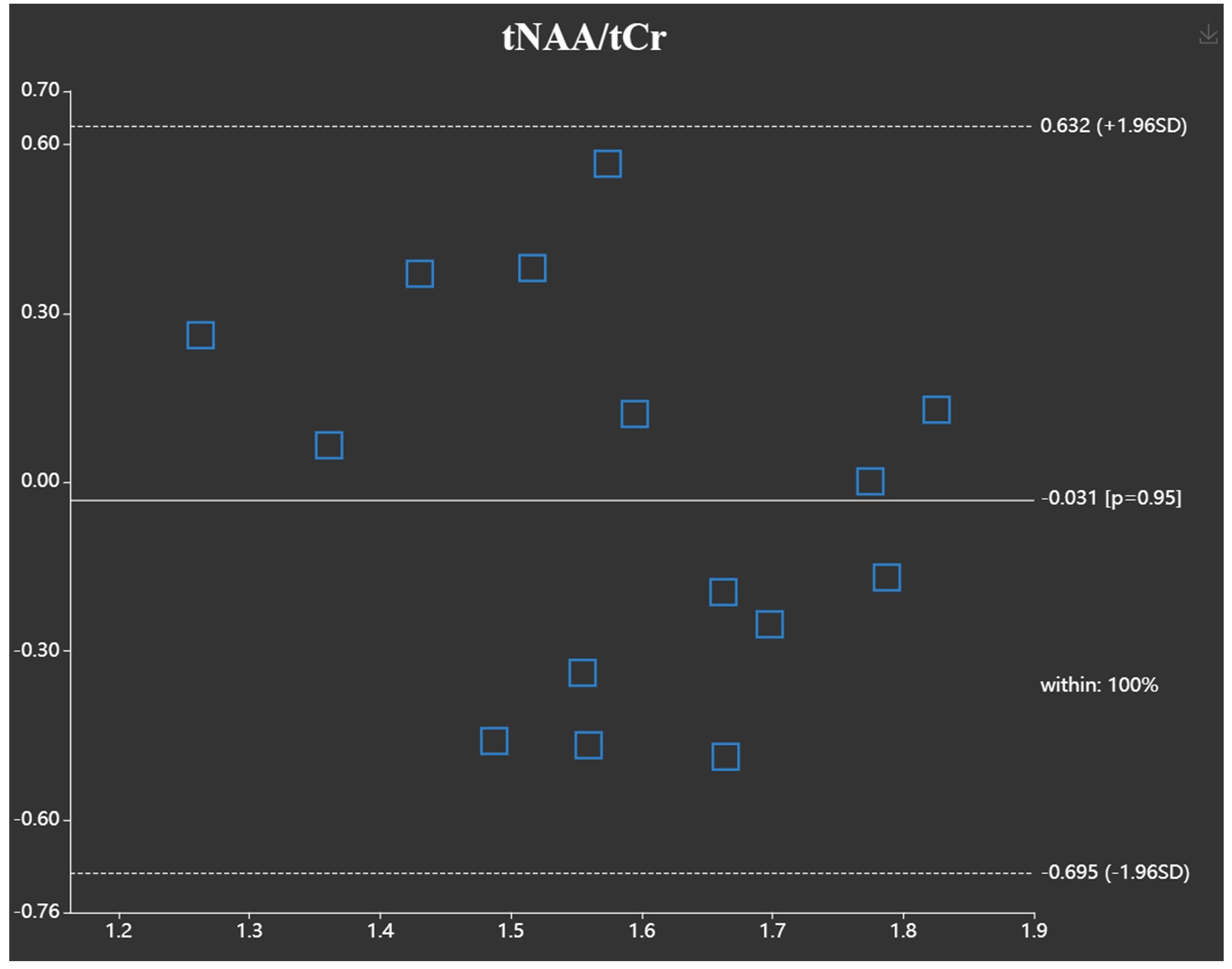}
    \caption{The Bland-Altman analysis for tNAA$/$tCr from 15 in vivo spectra of healthy volunteers. Each square represents the quantified result for each spectrum. The horizontal and vertical axes indicate the mean and difference, respectively, of the quantified results by the two quantification methods.}
    \label{FIG11}
\end{figure}

\subsection{Consistency analysis}
\label{Consistency analysis}
15 spectra (Philips RAW data) from healthy volunteers are uploaded to CloudBrain-MRS to quantify the consistency of QNet and LCModel.

\begin{figure}[t]
    \centering
    \includegraphics[width=0.48\textwidth,height=0.35\textwidth]{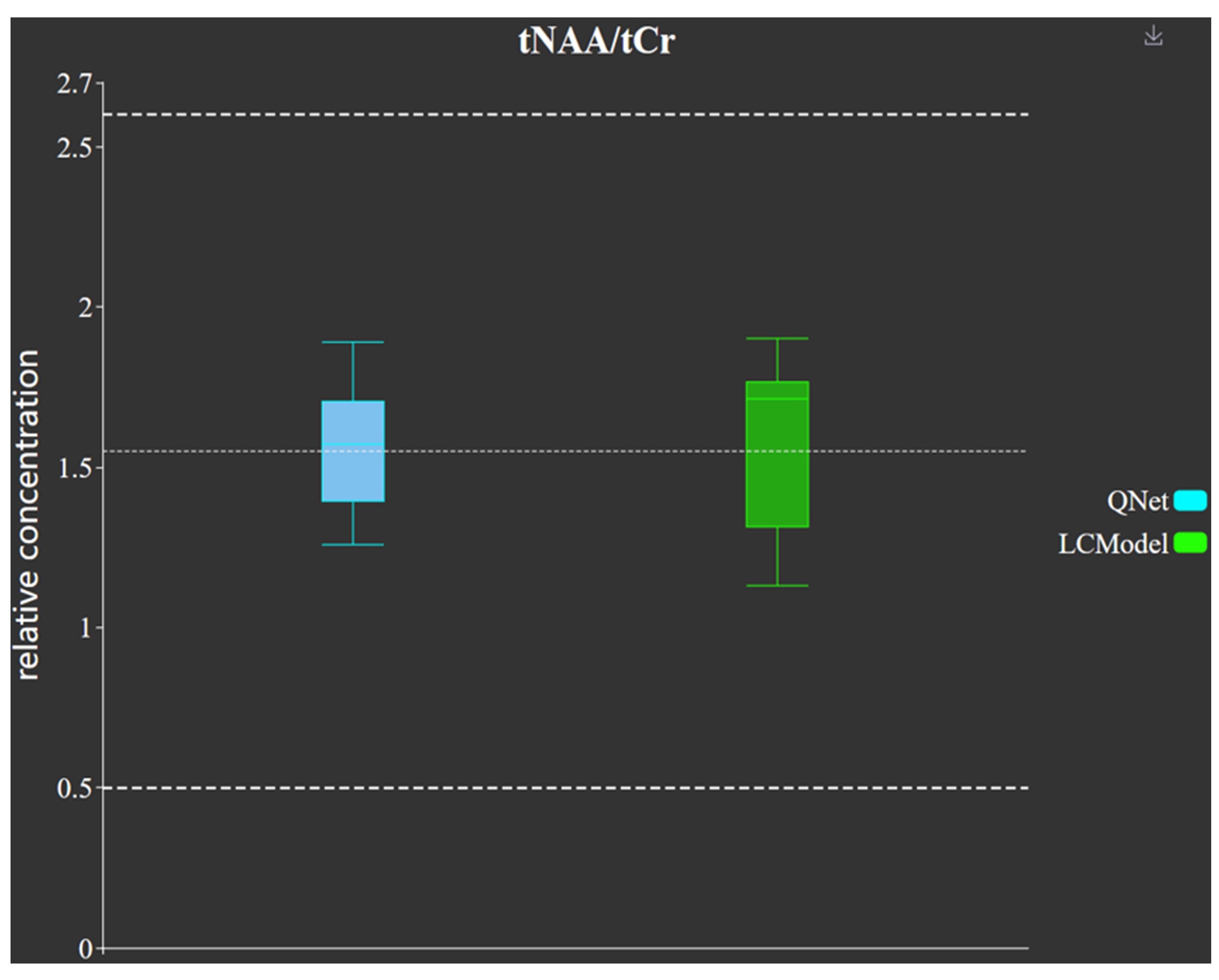}
    \caption{ Comparison of the tNAA$/$tCr ratio estimated by QNet and LCModel with box plots. The metabolite concentration range from the literature is marked with the upper and lower dashed lines and the mean 
value is indicated by the middle dashed line.}
    \label{FIG12}
\end{figure}

The Bland-Altman analysis of relative concentration tNAA$/$tCr is shown in Fig. \ref{FIG11}. The solid line represents the mean difference between the two methods on tNAA$/$tCr is 0.031. The p-value is the result of the independent t-test performed with the Standard Normal Distribution (SND) on the scatter plot, and the result is greater than 0.05, indicating no significant difference between the two methods on tNAA$/$tCr ratio. ±1.96SD (standard deviation) is used to represent the upper and lower limits of agreement, obtaining a 95\% confidence interval. In Fig. \ref{FIG11}, 96\% of the points fall within the confidence interval. These results suggest that QNet and LCModel have high consistency in the quantification results of tNAA$/$tCr for this spectrum. Additionally, Fig. \ref{FIG12} compares the box plots of tNAA$/$tCr ratio estimated by QNet and LCModel. For tNAA$/$tCr, both methods estimated the concentrations within the reasonable range \citep{lee2019intact,terpstra2016test,tkavc2009vivo,govindaraju2000proton,chen2023magnetic}, i.e. the distribution of concentrations is between the upper and lower dashed lines.

\section{Conclusion}
\label{Conclusion and future directions}
We have developed CloudBrain-MRS, a cloud computing platform that deploys artificial intelligence and classic algorithms to quantify MRS signals. Users can preprocess, quantify, and analyze MRS data in batches through an online browser without the need for environment installation or code compilation. CloudBrain-MRS is an open-access platform at
\url{https://csrc.xmu.edu.cn/CloudBrain.html}, and it also has been shared on \href{https://mrshub.org/}{MRSHub}, we will continue to do so for the next two years. For further improvement, CloudBrain-MRS will be validated with large-scale data and more useful algorithms will be deployed for more manufacturers and data types. To assist clinical research and diagnosis, we will enhance the analysis capabilities to generate examination reports using biomarkers and provide preliminary disease classification for reference by doctors. CloudBrain-MRS will be made into a quantify, and analyze MRS with a standard processing pipeline to serve the MRS research community. To verify the reliability of CloudBrain-MRS, more doctors and experts are expected to try it out.

\section*{Acknowledgments}
This work was partially supported by the National Natural Science Foundation of China (62122064, 61971361, 62331021, 62371410), the Natural Science Foundation of Fujian Province of China (2023J02005, 2021J011184), the President Fund of Xiamen University (20720220063), and Nanqiang Outstanding Talent Program of Xiamen University. The authors thank China Mobile for providing cloud computing services support. The authors thank Zhigang Wu, Liangjie Lin, and Jiazheng Wang from Philips and Jiayu Zhu and Xijing Zhang from United Imaging for technical support. The authors also thank Stephen W. Provencher for making LCModel public.










\bibliographystyle{elsarticle-num-names}

\bibliography{cas-dc-template}


\end{sloppypar}

\end{document}